\documentclass[12pt]{iopart}
\usepackage{iopams}

\usepackage{graphicx}

\begin{document} 

\title [A two-parameter random walk] {A two-parameter random walk with approximate exponential probability distribution}
\author {Erik Van der Straeten\footnote {Research Assistant of the Research Foundation - Flanders (FWO - Vlaanderen)}
and Jan Naudts}
\address {Departement Fysica, Universiteit Antwerpen,\\
          Universiteitsplein 1, 2610 Antwerpen, Belgium}
\eads {\mailto {Erik.VanderStraeten@ua.ac.be}, \mailto {Jan.Naudts@ua.ac.be}}

\date {}

\def\Io{{\mathbb I}}
\def\Zo{{\mathbb Z}}
\def\epsr{\epsilon_{\rm R}}
\def\epsl{\epsilon_{\rm L}}

\newtheorem{proposition}{Proposition}
\newtheorem{lemma}{Lemma}
\newenvironment{proof}
{\par\noindent {\bf Proof}}
{\par\strut\hfill$\square$\par\vskip 0.5cm}

\newcommand {\keywords}[1] {\noindent{\it Keywords\,\,}#1}

\newcommand{\be}{\begin{eqnarray}}
\newcommand{\ee}{\end{eqnarray}}

\renewcommand{\Im}{\,{\rm Im}\,}
\renewcommand{\Re}{\,{\rm Re}\,}


\begin {abstract}
We study a non-Markovian random walk in dimension 1. It depends on two parameters
$\epsr$ and $\epsl$, the probabilities to go straight on when walking to the right,
respectively to the left. The position $x$ of the walk after $n$ steps and the
number of reversals of direction $k$ are used to estimate $\epsr$ and $\epsl$.
We calculate the joint probability distribution $p_n(x,k)$ in closed form
and show that, approximately, it belongs to the exponential family.
\end {abstract}


\section {Introduction}

Consider a random walk starting in the origin $x=0$ of the lattice $\Zo$.
The probability that after $n$ steps the walk is in $x$ and changed
its direction $k$ times is denoted $p_n(x,k)$.
This paper investigates the question how $p_n(x,k)$ depends on model parameters.
We wonder whether it can be written into the form
\be
p_n(x,k)=D_n(x,k)\exp(G+\beta k + Fx).
\label {gibbs}
\ee
In this expression, $G$, $\beta$ and $F$ depend on model parameters.
However, the prefactor $D_n(x,k)$ does not depend on model parameters.
The function $\beta$ has the interpretation of an inverse temperature (in dimensionless units),
the function $F$ is an external force, the function $G$, when divided by $\beta$, is a free energy
and serves to normalize (\ref {gibbs}). 
A probability distribution $p_n(x,k)$ of the form (\ref {gibbs}) is said to
belong to the {\sl exponential family}. It has nice properties. In particular, averages
of $x$ and $k$ can be calculated by taking derivatives of $G$ with respect to the parameters.

Random walk models are omnipresent in statistical physics and have been studied extensively.
Quite often results are obtained in the limit of large $n$. Here, the focus is on all $n$.
Deviations from (\ref {gibbs}), found below, are neglegible in the large $n$-limit.
Standard techniques aim at calculating correlation functions. It is rather seldom
that exact expressions for probability distributions can be written down in closed form.
In the present model such closed form expression exists for $p_n(x,k)$, but probably not
for the marginals $p_n'(x)=\sum_kp_n(x,k)$ and $p_n'' (k)=\sum_xp_n(x,k)$.
Individual events have usualy such a small probability that they cannot be evaluated numerically.
In addition, in situations with a large number of degrees of freedom,
knowledge of $p_n(x,k)$ is not sufficient to evaluate moments of the distribution in closed form.
However, if a closed form expression of $p_n(x,k)$ is available then analytic relations can be used
to evaluate relevant quantities. 

The model, considered here, is that of a one-dimensional persistent random walk (see e.g.~\cite {GW02}) with drift.
Many generalizations of the persistent random walk can be found in literature, e.g.~for
continuous time \cite {MLW89}, or with a memory that goes back more than one step \cite {BL01}.
Planar persistent random walks have been studied in \cite {WS87, BCh04}.

Our model is a toy model that helps to understand features of more realistic models used in several
branches of physics. One such application, well-known since the pioneering work of Flory \cite {FP69},
is the use of random walks to model the geometry of polymers.
Persistent random walks play also a role in understanding the transition from
ballistic to diffusive transport \cite {BPM99,CG02,MS05}
and have been applied in financial physics,
see e.g.~\cite {KKK02}.

Recent technological progress
has made it possible to do experiments on single molecules and to measure the elongation
of a single polymer $\langle x\rangle$ as a function of applied force $F$
(see e.g.~\cite {SFB92,BMS94,KSB03}). The analysis of these experiments is based on the
assumption that $p_n'(x)$ is proportional to $\exp(-\beta V(x)+Fx)$ for some potential $V(x)$.
This relation follows from (\ref {gibbs}) with $\exp(-\beta V(x))=\sum_kD_n(x,k)e^{\beta k}$.
The latter expression shows that $V(x)$ is indeed a free energy, as claimed in \cite {KSB03}.
An in depth discussion of these experiments based on the results of the present paper is found in \cite {VN06}.

In the next section the model is introduced. In Section 3
average values for position $x$ and number of reversals $k$ are
calculated using the method of generating functions.
In Section 4 the number of walks ending in $x$ after $n$ steps,
and having a given number of reversals $k$, is calculated.
These counting results are used in Section 5
to write down the joint probability distribution $p_n(x,k)$.
Section 6 considers the dependence of $p_n(x,k)$
on the parameters $\epsr$ and $\epsl$ and tries to answer the question 
whether this two-parameter probability distribution function (pdf) belongs to the exponential family.
Section 7 shows how to calculate averages starting from the
knowledge that the pdf is exponential.
The final section gives a short discussion of the results.

\section{Model}

Consider a discrete-time random walk on the one-dimensional lattice $\Zo$. The probability
of the walk to step to the right (i.e., with increasing position)
equals $\epsr$ when coming from the left
and $1-\epsl$ when coming from the right. This is not a Markov chain since the walk
remembers the direction it comes from.
Let $x_n$ be the position of the walk after $n$ steps.
Let $\sigma_n=x_n-x_{n-1}$ be the direction of the $n$-th step.
Then $x_{n+1}=x_n+1$ with probability
\be
\frac 12(1+\sigma_n)\epsr+\frac 12(1-\sigma_n)(1-\epsl),
\ee
$x_{n+1}=x_n-1$ otherwise.
The process of the increments $\sigma_n$ is a two-state Markov chain
with transition matrix
\be
P=\left(\begin {array}{lr}
\epsr &1-\epsr\\
1-\epsl    &\epsl
\end{array}\right).
\ee
In the stationary state $\sigma_n$ equals $\pm 1$ with probability $p_\pm^{(0)}$ given by
\be
p_+^{(0)}=\frac {1-\epsl}{2-\epsr-\epsl},
\qquad
p_-^{(0)}=\frac {1-\epsr}{2-\epsr-\epsl}.
\label {stationary}
\ee

Let $k_n$ denote the number of reversals of the walk after $n$ steps.
By definition a reversal occurs at step $n$ if $\sigma_{n-1}\sigma_n=-1$.
Hence one has
\be
x_n&=&\sum_{j=1}^n\sigma_j\\
k_n&=&\frac 12\sum_{j=1}^n(1-\sigma_{j-1}\sigma_j).
\ee
The quantity of interest in this paper is the joint probability of
position $x_n$ and number of reversals $k_n$. The obvious initial conditions
are $x_0=0$ and $\sigma_0=\pm 1$ with probability $p_\pm^{(0)}$.

The physical interpretation of the model is twofold. The random walk is a simple model of
a polymer with $n$ units. Energy is proportional to minus the number of
reversals $k_n$. The position of the end point $x_n$ measures the
effect of an external force applied to the end point.
Alternatively, $k_n$ is the number of domains (decreased by 1 if $\sigma_1=\sigma_0$)
of an Ising chain, and $x_n$ is the total magnetization. Indeed,
the variables $\sigma_n$ describe Ising spins on a one-dimensional lattice.
A domain is then a set of subsequent sites where the spins all have the
same value, either up (+1) or down (-1). The boundary between two domains
involves a reversal ($\sigma_{j-1}\sigma_j=-1$).

\section {Generating functions}

Let $p^\pm(n,x,k)$ denote the probability that $\sigma_n=\pm 1$, $x_n=x$, and $k_n=k$.
The joint proability distribution, searched for, is then
\be
p_n(x,k)=p_n^+(x,k)+p_n^-(x,k).
\ee
The following recursion relations hold
\be
p_n^+(x,k)&=&\epsr p_{n-1}^+(x-1,k)+(1-\epsl) p_{n-1}^+(x-1,k-1)\\
p_n^-(x,k)&=&(1-\epsr) p_{n-1}^+(x+1,k-1)+\epsl p_{n-1}^+(x+1,k).
\ee
Introduce generating functions
\be
f_\pm^{(n)}(w,z)&=&\sum_{x=-n}^n w^x\sum_{k=0}^{n}z^kp^\pm_n(x,k).
\ee
and a similar expression for $f^{(n)}(w,z)$.
They satisfy
\be
\left(\begin {array}{c}
f_+{(n)}\\
f_-{(n)}
      \end{array}\right)
=M(w,z)
\left(\begin {array}{c}
f_+{(n-1)}\\
f_-{(n-1)}
      \end{array}\right)
\ee
with
\be
M(w,z)=
\left(\begin {array}{lr}
\epsr w &(1-\epsl)wz\\
(1-\epsr)z/w &\epsl/w
      \end {array}\right).
\ee
It is possible to calculate the $n$-th power of this matrix by first diagonalizing it.
The result is
\be
M^n(w,z)
&=&\frac 12(\lambda_+^n+\lambda_-^n)
\left(\begin {array}{lr}
1 & 0\\
0 & 1 \end {array}\right)\cr
& &
+\frac 1{2\nu}(\lambda_+^n-\lambda_-^n)
\left(\begin {array}{lr}
w\epsr-\epsl/w &2(1-\epsl)wz\\
2(1-\epsr)z/w    &-w\epsr+\epsl/w
      \end {array}\right),
\ee
with
\be
\lambda_\pm=\frac 12(\epsr w+\epsl/w\pm \nu)
\ee
and
\be
\nu=\sqrt{(\epsr w-\epsl/w)^2+4(1-\epsr)(1-\epsl)z^2}.
\ee

Let us now consider initial values. Note that
$f_\pm^{(0)}(w,z)$ is not yet defined because $k_0$ involves $\sigma_{-1}$,
which is undetermined. Starting point is therefore $f_\pm^{(1)}(w,z)$,
which is found to be given by
\be
f_\pm^{(1)}(w,z)=M(w,z)
\left(\begin {array}{c}
p_+^{(0)}\\p_-^{(0)}
\end {array}\right).
\ee
Hence, it is obvious to define
\be
f_\pm^{(0)}(w,z)=
\left(\begin {array}{c}
p_+^{(0)}\\p_-^{(0)}
\end {array}\right).
\ee
The generating function $f^{(n)}(w,z)$ is now explicitly known
\be
f^{(n)}(w,z)
&=&\frac 12(\lambda_+^n+\lambda_-^n)
+\frac 1{2\nu}(\lambda_+^n-\lambda_-^n)
\bigg[(\epsr w-\epsl/w)(p_+^{(0)}-p_-^{(0)})\cr
& &\qquad +2(1-\epsr)p_+^{(0)}z/w+2(1-\epsl)p_-^{(0)}wz\bigg].\cr
& &
\ee
It can be used to calculate expectation values by taking derivatives. For example,
\be
\langle k_n\rangle&=&\frac {\partial\,}{\partial z}\bigg|_{w=z=1}f^{(n)}(w,z)\cr
&=&2n\frac {(1-\epsr)(1-\epsl)}{2-\epsr-\epsl},
\label {kav}
\ee
and
\be
\langle x_n\rangle&=&\frac {\partial\,}{\partial w}\bigg|_{w=z=1}f^{(n)}(w,z)\cr
&=&n\frac {\epsr-\epsl}{2-\epsr-\epsl}.
\label {xav}
\ee

\section {Counting walks}

The present section is temporarily limited to the special case $\epsr=\epsl=1/2$.
From the next section on the general model will be considered again.
Indeed, we first determine the number of walks $c_\pm(n,x,s)$ which,
starting in the origin in direction $\pm 1$, end in $x$ after $n$ steps
and have $s$ segments. The result does not depend on the value of 
$\epsr$ and $\epsl$. Hence the calculation can be done in the simplest case.
In the next section the result will be used to calculate the joint probability distribution $p_n(x,k)$
for the general model.

Divide the walk into segments of constant $\sigma_j$. Number these segments from
1 to  $s_n$.
Note that $s_n=k_n+1$ if $\sigma_1=\sigma_0$, $s_n=k_n$ otherwise. This means that the number
of segments equals 1 plus the number of reversals, not counting
the initial reversal at $x=0$, if present.
Let $\tau_j$ denote the length of the $j$-th segment. The probability that
segment $j$ has length $l$ equals $2^{-l}$. The probability of counting  $s$ segments
in a walk of $n$ steps satisfies
\be
{\cal P}(s_n=s)
&=&{\cal P}\bigg(\sum_{j=1}^{s-1}\tau_j<n\le \sum_{j=1}^s\tau_j\bigg)\cr
&=&\sum_{l_1=1}^\infty\cdots \sum_{l_s=1}^\infty 2^{-l_1-\cdots-l_s}
\Io\{l_1+\cdots+l_{s-1} < n \le l_1+\cdots+l_s \}\cr
&=&\sum_{m=s-1}^{n-1}2^{-m}
\sum_{l_1=1}^\infty\cdots \sum_{l_{s-1}=1}^\infty\delta_{m,l_1+\cdots+l_{s-1}}
\sum_{l_s=n-m}^\infty 2^{-l_s}\cr
&=&2^{-(n-1)}\sum_{m=s-1}^{n-1}
\sum_{l_1=1}^\infty\cdots \sum_{l_{s-1}=1}^\infty\delta_{m,l_1+\cdots+l_{s-1}}\cr
&=&2^{-(n-1)} {n-1 \choose s-1}.
\ee
Hence the conditional probability given a certain number of segments equals
\be
{\cal P}(\tau_1=l_1,\cdots,\tau_ {s-1}=l_{s-1}\,|\,s_n=s)
&=& {n-1 \choose s-1}^{-1}.
\ee
This means that the variables $\tau_j$, after conditioning on a given number of segments,
become uniformely distributed. This observation simplifies the following
calculation.

The position $x_n$ of the walk after $n$ steps, assuming $s_n$ segments,
can be expressed into the segment lengths as
\be
x_n&=&\sigma_1\bigg(\tau_1-\tau_2+\cdots\pm\tau_{s_n-1}\mp(n-\sum_{j=1}^{s_n-1}\tau_j)\bigg).
\ee
The $\pm$-sign depends on whether the number of segments $s_n$ is even or odd
and equals $(-1)^{s_n}$. One obtains
\be
x_n&=&\sigma_1\bigg(2(\tau_1+\tau_3+\cdots+\tau_{s_n-1})-n\bigg)
\qquad\hbox{ if }s_n\hbox{ is even}\cr
&=&\sigma_1\bigg(n-2(\tau_2+\tau_4+\cdots+\tau_{s_n-1})\bigg)
\qquad\hbox{ if }s_n\hbox{ is odd}.
\ee

For simplicity let us first consider the case of an even number of segments.
Let $s>0$ be even. Then one has
\be
& &{\cal P}(x_n=x,s_n=s)\cr
&=&{\cal P}(s_n=s)
{\cal P}(\tau_1+\tau_3\cdots+\tau_{s-1}=(n+\sigma_1x)/2
\,|\,s_n=s)\cr
&=&2^{-(n-1)}\sum_{l_1=1}^\infty\cdots\sum_{l_{s-1}=1}^\infty
\Io\{\sum_{j=1}^{s-1}l_j<n,l_1+l_3+\cdots +l_{s-1}=(n+\sigma_1 x)/2\}\cr
&=&2^{-(n-1)}\sum_{l_1=1}^\infty\sum_{l_3=1}^\infty\cdots\sum_{l_{s-1}=1}^\infty
\Io\{ l_1+l_3+\cdots +l_{s-1}=(n+\sigma_1 x)/2\}\cr
& &\times
\sum_{l_2=1}^\infty\sum_{l_4=1}^\infty\cdots\sum_{l_{s-2}=1}^\infty
\Io\{l_2+l_4+\cdots+l_{s-2}<(n-\sigma_1x)/2\}\cr
&=&2^{-(n-1)}\Theta_n^{\sigma_1}(x,s)C(n+x-2,s-2)C(n-x-2,s-2)
\ee
with
\be
C(n,m)=\frac {n!!}{m!!((n-m)!!}.
\ee
and with $\Theta_n^\pm(x,s)$ equal 1 if there exists a walk of $n$ steps,
starting in the origin in direction $\pm 1$, ending in $x$,
and containing $s$ segments, and zero otherwise.

If $s$ is odd, $s\ge 3$, then one has
\be
& &{\cal P}(x_n=x,s_n=s)\cr
&=&{\cal P}(s_n=s)
{\cal P}(\tau_2+\tau_4\cdots+\tau_{s-1}=(n-\sigma_1x)/2
\,|\,s_n=s)\cr
&=&2^{-(n-1)}\sum_{l_1=1}^\infty\cdots\sum_{l_{s-1}=1}^\infty\cr
& &\times
\Io\{\sum_{j=1}^{s-1}l_j<n,l_2+l_4+\cdots +l_{s-1}=(n-\sigma_1 x)/2\}\cr
&=&2^{-(n-1)}\sum_{l_1=1}^\infty\sum_{l_3=1}^\infty\cdots\sum_{l_{s-2}=1}^\infty\cr
& &\times
\Io\{ l_1+l_3+\cdots +l_{s-2}<(n+\sigma_1 x)/2\}\cr
& &\times
\sum_{l_2=1}^\infty\sum_{l_4=1}^\infty\cdots\sum_{l_{s-1}=1}^\infty
\Io\{l_2+l_4+\cdots+l_{s-1}=(n-\sigma_1x)/2\}\cr
&=&2^{-(n-1)}\Theta_n^{\sigma_1}(x,s)C(n+\sigma_1x-2,s-1)C(n-\sigma_1x-2,s-3).\cr
& &
\ee
Note that, in case of an odd number of segments, the number of walks
ending in $x_n$ depends on whether the walk starts to the left or to the right.

Finally, if $s=1$ then there is clearly only one walk ending in the point $x_n$.

\section {The joint probability distribution}

Let us return to the general case with arbitrary $\epsr$ and $\epsl$.
The probability of a given $n$-step walk depends only on $\sigma_0$,
and on the final values $x_n$ and $k_n$. To see this, note that a segment of length $\tau$
has probability $(1-\epsr)\epsr^{\tau-1}$ if the direction
is positive, and $(1-\epsl)\epsl^{\tau-1}$ if the direction is
negative. A factor $\epsr$ can be associated with every step to the right,
and $\epsl$ with every step to the left. But then a factor $(1-\epsl)/\epsr$,
respectively $(1-\epsr)/\epsl$,
must be associated with every reversal of direction from leftgoing to rightgoing,
respectively rightgoing to leftgoing.

The number of steps to the right respectively to the left is $(n+x_n)/2$,
respectively $(n-x_n)/2$. The number of reversals from leftgoing to rightgoing
is denoted $k_n^-$, from rightgoing to leftgoing $k_n^+$. They depend on whether the number of reversals
is even or odd. If $k_n$ is odd then
\be
k_n^\pm&=&(k_n\pm\sigma_0)/2.
\ee
Obviously is $k_n^-+k_n^+=k_n$ and $k_n^+-k_n^-=\sigma_0$.
On the other hand, if $k_n$ is even then the number of reversals is
$k_n^-=k_n^+=k_n/2$, independent of the direction $\sigma_0$.
In both cases, the probability of the $n$-step walk, given that it ends
in $x$, has $k^+$ reversals when going right and $k^-$ when going left, is
\be
& &
\gamma_n(x,k^+,k^-)\equiv
\epsr^{\frac {n+x}2}\epsl^{\frac {n-x}2}
\left(\frac {1-\epsr}\epsl\right)^{k^+}
\left(\frac {1-\epsl}\epsr\right)^{k^-}
.\cr
& &
\ee
The number of such walks is denoted $D_n^{\sigma_0}(x,k^+,k^-)$ and equals
\be
D_n^{\sigma_0}(x,k^+,k^-)&=&
\Theta_{n+1}^{\sigma_0}(x+\sigma_0,k^++k^-+1)\Xi^{\sigma_0}(k^+,k^-)\cr
& &\times
C(n-1+x+\sigma_0,2k^-+\sigma_0-1)\cr
& &\times
C(n-1-x-\sigma_0,2k^+-\sigma_0-1).\cr
& &
\label {countres}
\ee
The function $\Xi^{\sigma_0}(k^+,k^-)$ equals 1 if $k^+=k^-$ or $k^+-k^-=\sigma_0$
and zero otherwise.
To see from where (\ref {countres}) follows, consider a walk of $n+1$ steps starting at position $-\sigma_0$
and apply the results of the previous section. Note that the number of segments
of this walk is $k^++k^-+1$.

The final result for the joint probability distribution $p_n(x,k)$ is then
\be
p_n(x,k)&=&p_+^{(0)}q_n^+(x,k)+p_-^{(0)}q_n^-(x,k)
\label {pfinal}
\ee
with the probability distribution $q_n^\pm (x,k)$
given by
\be
q_n^\pm(x,k)
&=&D_n^\pm\left(x,\frac k2,\frac k2\right)\gamma_n\left(x,\frac k2,\frac k2\right),
\qquad\qquad k\hbox { even}\cr
&=&D_n^\pm\left(x,\frac {k\pm 1}2,\frac {k\mp 1}2\right)\gamma_n\left(x,\frac {k\pm 1}2,\frac {k\mp 1}2\right),
\qquad k\hbox{ odd}.\cr
& &
\label {qpdf}
\ee

As an example let us calculate
\be
p_4(2,3)&=&p_+^{(0)}D_4^+(2,2,1)\gamma_4(2,2,1)+p_-^{(0)}D_4^-(2,1,2)\gamma_4(2,1,2).
\ee
One has
\be
D^+_4(2,2,1)&=&\Theta^+_5(3,4)\Xi^+(2,1)C(6,2)C(0,2)\\
D^-_4(2,1,2)&=&\Theta^-_5(1,4)\Xi^-(1,2)C(4,2)C(2,2).
\ee
Clearly, $\Theta^+_5(3,4)=0$ and $\Theta^-_5(1,4)=1$.
Using
\be
\gamma_4(2,1,2)=\epsr(1-\epsr)(1-\epsl)^2
\ee
one obtains
\be
p_4(2,3)=2\epsr\frac {(1-\epsr)^2(1-\epsl)^2}{2-\epsr-\epsl}.
\ee
This example shows that sometimes only one of the
two terms contributing to the r.h.s.~of (\ref {pfinal}) does not vanish.

\section {Exponential family}

One can write
\be
\gamma_n\left(x,\frac k2,\frac k2\right)=
\exp(G+\beta k+Fx)
\ee
with
\be
F &=&\frac 12\ln\frac\epsr\epsl
\label {Fres}\\
\beta &=&-\frac 12\ln\frac {\epsr\epsl}{(1-\epsr)(1-\epsl)}
\label {betares}\\
G&=&\frac {n}2\ln\epsr\epsl.
\label {Gres}
\ee
This allows us to write $q_n^\pm(x,k)$, appearing in our main result (\ref {pfinal}), as
\be
q_n^\pm(x,k)&=&
D_n^\pm\left(x,\frac {k\pm\Delta}2,\frac {k\mp\Delta}2\right)
\exp(G+\beta k+Fx\pm\gamma\Delta),
\label {qexact}
\ee
where
\be
\gamma=\frac 12\ln\frac {(1-\epsr)\epsr}{(1-\epsl)\epsl}
\label {gammares},
\ee
and with $\Delta=1$ if $k$ is odd, and zero if $k$ is even.
Hence the probability distributions $q_n^\pm(x,k)$ belong to the exponential family,
however not with two but with three parameters $\beta$, $F$, and $\gamma$.
The third parameter $\gamma$ controls boundary effects.
Hence, $p_n(x,k)$ is a superposition of two distributions
$q_n^\pm(x,k)$, both belonging to the exponential family. However, the domains on which
these two pdfs differ from zero are not identical.

If $n$ is large then the variable $\Delta$ can usually be neglected, being small compared to
typical values of $k$ and $x$.
One obtains the approximate result that, for those values of $x$ and $k$ for which $p_n(x,k)\not=0$,
\be
p_n(x,k)\simeq {\frac {n+x}2\choose\frac k2} {\frac {n-x}2\choose\frac k2}
\exp(G+\beta k+Fx).
\label {approxgibbs}
\ee
This shows that $p_n(x,k)$ approximately belongs to the exponential family with
two parameters $\beta$ and $F$.
Deviations between l.h.s.~and r.h.s.~of (\ref  {approxgibbs}) occur for two reasons: there is a subtle difference in expressions for
even $k$ and for odd $k$, and there is a small dependence on the initial condition $\sigma_0$.

\section {Calculating averages}

Let us now see what exponential expressions are good for.
First consider the approximate expression (\ref {approxgibbs}).
From $\sum_{x,k}p_n(x,k)=1$ follows
\be
0&\simeq&\sum_n\frac {\partial\,}{\partial\beta}p_n(x,k)=\frac {\partial G}{\partial\beta}+\langle k_n\rangle\\
0&\simeq&\sum_n\frac {\partial\,}{\partial F}p_n(x,k)=\frac {\partial G}{\partial F}+\langle x_n\rangle.
\ee
Using (\ref {Gres}) there follows
\be
0&=&\frac n{2\epsr}-\frac 1{2\epsr(1-\epsr)}\langle k_n\rangle+\frac 1{2\epsr}\langle x_n\rangle\\
0&=&\frac n{2\epsl}-\frac 1{2\epsl(1-\epsl)}\langle k_n\rangle-\frac 1{2\epsl}\langle x_n\rangle.
\ee
When solving these equations for $\langle k_n\rangle$ and $\langle x_n\rangle$ one recovers
(\ref {kav}, \ref {xav}). Hence, from the approximate result (\ref {approxgibbs}),
which one can guess whithout hard work, one obtains immediately exact results for the averages
$\langle k_n\rangle$ and $\langle x_n\rangle$.

Let us now try to do the same starting from the exact expressions (\ref {pfinal}, \ref {qexact}). 
From the normalization of $q_n^\pm(x,k)$ follows the set of equations
\be
0&=&\frac n{2\epsr}-\frac 1{2\epsr(1-\epsr)}\langle k_n\rangle^\pm+\frac 1{2\epsr}\langle x_n\rangle^\pm
\pm\frac {1-2\epsr}{2\epsr(1-\epsr)}\langle\Delta\rangle^\pm\\
0&=&\frac n{2\epsl}-\frac 1{2\epsl(1-\epsl)}\langle k_n\rangle^\pm-\frac 1{2\epsl}\langle x_n\rangle^\pm
\mp\frac {1-2\epsl}{2\epsl(1-\epsl)}\langle\Delta\rangle^\pm.
\ee
They can be written as
\be
\langle x_n\rangle^\pm&=&n\frac {\epsr-\epsl}{2-\epsr-\epsl}\mp2\frac {1-\epsr-\epsl}{2-\epsr-\epsl}\langle\Delta\rangle^\pm
\label {xpm}\\
\langle k_n\rangle^\pm&=&2n\frac {(1-\epsr)(1-\epsl)}{2-\epsr-\epsl}
\mp\frac {\epsr-\epsl}{2-\epsr-\epsl}\langle\Delta\rangle^\pm.
\label {kpm}
\ee
For large $n$, the effect of the terms in $\Delta$ is small, as can be seen from these equations.

Comparison with (\ref {kav}, \ref {xav}) gives
\be
p_+^{(0)}\langle\Delta\rangle^+=p_-^{(0)}\langle\Delta\rangle^-.
\ee
Hence one can write
\be
\langle\Delta\rangle^+=\frac {2-\epsr-\epsl}{2(1-\epsl)}\langle\Delta\rangle\\
\langle\Delta\rangle^-=\frac {2-\epsr-\epsl}{2(1-\epsr)}\langle\Delta\rangle.
\ee
It is however not clear how to obtain a closed form expression for $\langle\Delta\rangle$,
which is the probability that $k_n$ is odd.

The averages (\ref {xpm}, \ref {kpm}) are calculated with boundary conditions
$\sigma_0=+1$ or $\sigma_0=-1$. The expressions are the sum of a part
independent of the boundary condition and a small contribution
which depends on the boundary condition and on the probability
$\langle\Delta\rangle$ that the number of reversals $k_n$ is odd.
An annoying consequence of the fact that the probabilities $q_n^\pm(x,k)$
belong to the exponential family with three parameters instead of two
is that the averages of $x_n$ and $k_n$ cannot be obtained from (\ref {qexact})
by simple taking of derivatives. Even when the results of Section 3
are invoqued, not all quantities can be determined. In particular the 
probability $\langle\Delta\rangle$ cannot be obtained in this way.

Simple random walk corresponds with the choice $\epsr=\epsl=1/2$.
This implies infinite temperature (i.e.~vanishing $\beta$) and absence of drift ($F=0$).
The third parameter $\gamma$ vanishes as well. Also random walk with drift
is a special case, corresponding with $\epsr+\epsl=1$. Again, $\beta=0$
and $\gamma=0$ follow. A persistent random walk is obtained when $\epsr=\epsl$.
This implies $F=0$, but non-vanishing $\beta$ and $\gamma$.

\section {Discussion}

We have studied a simple model of random walk depending on two parameters
$\epsr$ and $\epsl$. The parameters are estimated using the position $x_n$
of the walk after $n$ steps, and the number of reversals of direction
$k_n$. The technique of generating functions is used to calculate
averages $\langle x_n\rangle$ and $\langle k_n\rangle$. Next, 
explicit expressions are obtained for the number of walks that end in the
same position $x_n$ and have the same number of reversals $k_n$. These
counts are used to write an explicit result (\ref {pfinal})
for the joint probability distribution $p_n(x,k)$.
In the final part of the paper we try to write this joint probability distribution
in the form of an exponential family. This succeeds only in an
approximate manner. The distribution $p_n(x,k)$ is a superposition
of two pdfs $q_n(x,k)$, both belonging to the exponential family,
but with three parameters instead of two. The third parameter controls the probability
that the number of reversals is odd. The difference between walks
with even or odd number of reversals vanishes in the limit of large $n$.

\begin{figure}
\centerline {\includegraphics[width=8cm] {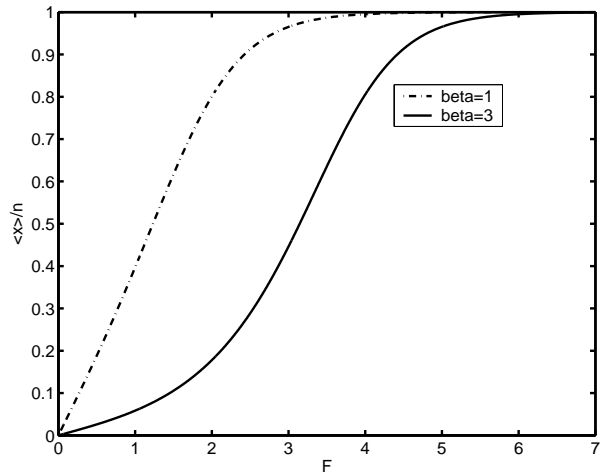}}
{\caption
{Average position $\langle x\rangle$, divide by $n$, as a function of external force $F$ for two different
values of $\beta$; based on Equations (\ref {xav}, \ref {Fres}, \ref {betares}).
}
\label{fig1}
}
\end{figure}

Some of the main results of the paper are explicit expressions (\ref {Fres}, \ref {betares})
for thermodynamic parameters $\beta$ and $F$ in terms of the model parameters $\epsr$ and $\epsl$.
They are used in Figure 1 to plot average position as a function of external force $F$.
The latter quantity can be measured experimentally. Our result shows a typical sigmoidal curve,
in  qualitative agreement with the measurements of \cite {SFB92}.
A more profound discussion of the implications of the present work is found in \cite {VN06}.
 
The calculation of the probability $\langle \Delta\rangle$ that the number of reversals
is odd is an open problem. Also did we not succeed to obtain a closed expression for the marginal
distribution $p_n'(x)=\sum_kp_n(x,k)$ of the position of the walker.
Note that the result of Section 4, counting walks with given $x$ and $k$,
is not needed in later sections to derive expressions for average position and average
number of reversals. This is a positive consequence of knowing that the parameter dependence
of the probability distributions of (\ref {qexact}) is exponential.
There is good hope to find distributions belonging to the exponential family also in more general
models, because exact relations can be derived, even in cases where counting walks would
raise an unsurmountable problem.

Deviations from exponential distribution, as in expression (\ref {approxgibbs}), are due to memory effects.
The walker remembers initial conditions, even if these are carefully chosen.
Reason here is that the process is non-markovian. Of course, these effects are negligible when
the number of steps $n$ is large. In many realistic models long range interactions
produce memory effects which remain important for large $n$. E.g., in polymers
the excluded volume effect causes long range interactions.
Such models are less suited for rigorous analysis.
We expect that deviations from exponential dependence, found for finite $n$ in the present model,
will occur in models with long range interactions, even in the limit of large system size.

\ack
We thank Frank den Hollander for suggesting the techniques used in Section 4,
and for his interest in the present work.


\end {document}